\documentclass[preprint,prd,aps,showpacs,showkeys,nofootinbib]{revtex4}
\usepackage{graphicx}
\usepackage{float}
\begin{document}

\title{$\bar{B}\rightarrow X_s\gamma$ in the $\mu\nu$SSM}

\author{
Hai-Bin Zhang$^{a,b,}$\footnote{hbzhang@mail.dlut.edu.cn},
Guo-Hui Luo$^{a,}$\footnote{ghuiluo@qq.com},
Tai-Fu Feng$^{a,b,}$\footnote{fengtf@hbu.edu.cn},
Shu-Min Zhao$^{b}$,
Tie-Jun Gao$^{c}$,
Ke-Sheng Sun$^{a}$}

\affiliation{$^a$Department of Physics, Dalian University of Technology, Dalian, 116024, China\\
$^b$Department of Physics, Hebei University, Baoding, 071002, China\\
$^c$Institute of theoretical Physics, Chinese Academy of Sciences, Beijing 100190, China}


\begin{abstract}
The $\mu\nu$SSM, one of supersymmetric extensions beyond the Standard Model, introduces three singlet right-handed neutrino superfields to solve the $\mu$ problem and can generate three tiny Majorana neutrino masses through the seesaw mechanism. In this work, we investigate the rare decay process $\bar{B}\rightarrow X_s\gamma$ in the $\mu\nu$SSM, under a minimal flavor violating assumption for the soft breaking terms. Constrained by the SM-like Higgs with mass around 125 GeV, the numerical results show that the new physics can fit the experimental data for $\bar{B}\rightarrow X_s\gamma$ and further constrain the parameter space.
\end{abstract}


\maketitle

\section{Introduction\label{sec1}}

The rare decay $\bar{B}\rightarrow X_s\gamma$ is one of the most promising windows to detect the new physics (NP) beyond the Standard Model (SM), since the theoretical evaluation on the decay width of the channel is induced by loop diagrams which are sensitive to the new fields coupled to bottom quark. The current combined  experimental data for the branching ratio of $\bar{B}\rightarrow X_s\gamma$ measured by CLEO \cite{refS.C}, BELLE \cite{refK.A,refA.L} and BABAR \cite{refJ.P,refJ.P1,refJ.PT,refB.A} give \cite{refS.S}
\begin{eqnarray}
{\rm{Br}}(\bar{B}\rightarrow X_s\gamma)=(3.37\pm0.23)\times10^{-4}.
\end{eqnarray}
Up to the next-next-to-leading order (NNLO), the theoretical prediction of ${\rm{Br}}(\bar{B}\rightarrow X_s\gamma)$ in the SM reads~\cite{refmm,ref18,refK.C,refC.G,refK.A1,refA.A}
\begin{eqnarray}
{\rm{Br}}(\bar{B} \rightarrow X_s\gamma) = (3.15 \pm0.23)\times10^{-4},
\end{eqnarray}
which coincides with the experimental result very well.

As a supersymmetric extension of the SM, the $\mu$ from $\nu$ Supersymmetric Standard Model ($\mu\nu$SSM) \cite{ref2,ref3,ref4} solves the $\mu$ problem \cite{ref5} of the Minimal Supersymmetric Standard Model (MSSM) \cite{ref6,ref7,ref8} through the lepton number breaking couplings between the right-handed neutrino superfields and the Higgses $\epsilon _{ab}{\lambda _i}\hat \nu _i^c\hat H_d^a\hat H_u^b$ in the superpotential.  The $\mu$ term is generated spontaneously through right-handed neutrino superfields vacuum expectation values (VEVs), $\mu  = {\lambda _i}\left\langle {\tilde \nu _i^c} \right\rangle$, once the electroweak symmetry is broken (EWSB). In this paper, we analyze the flavor changing neutral current (FCNC) process $\bar{B}\rightarrow X_s\gamma$ within the framework of the $\mu\nu$SSM under a minimal flavor violating version
for the soft breaking terms, constrained by the SM-like Higgs with mass around 125 GeV.

This paper has the following structure. In Section~\ref{sec2}, we present the $\mu\nu$SSM briefly, including its superpotential and the general soft SUSY-breaking terms. Section~\ref{sec3} contains the effective Lagrangian method and our notations. Then we get the Wilson coefficients of the process $\bar{B}\rightarrow X_s\gamma$. In Section~\ref{sec4}, we give the numerical analysis, under some assumptions and constraints on parameter space. The conclusion is given in Section~\ref{sec5}. Some formulae are collected in Appendixes~\ref{appendix-L}--\ref{appendix-integral}.

\section{{The $\mu\nu$SSM}\label{sec2}}

Besides the superfields of the MSSM, the $\mu\nu$SSM introduces three exotic right-handed neutrino superfields $\hat{\nu}_i^c$, $(i=1,\;2,\;3)$, which have nonzero VEVs. The corresponding superpotential of the $\mu\nu$SSM is given by~\cite{ref2}
\begin{eqnarray}
&&W ={\epsilon _{ab}} ({Y_{{u_{ij}}}}\hat H_u^b\hat Q_i^a\hat u_j^c + {Y_{{d_{ij}}}}\hat H_d^a\hat Q_i^b\hat d_j^c
+ {Y_{{e_{ij}}}}\hat H_d^a\hat L_i^b\hat e_j^c \nonumber \\
&&\qquad + {Y_{{\nu _{ij}}}}\hat H_u^b\hat L_i^a\hat \nu _j^c ) -  {\epsilon _{ab}}{\lambda _i}\hat \nu _i^c\hat H_d^a\hat H_u^b + \frac{1}{3}{\kappa _{ijk}}\hat \nu _i^c\hat \nu _j^c\hat \nu _k^c\;,
\label{super-w}
\end{eqnarray}
where $\hat H_d^T = \Big( {\hat H_d^0,\hat H_d^ - } \Big)$, $\hat H_u^T = \Big( {\hat H_u^ + ,\hat H_u^0} \Big)$, $\hat Q_i^T = \Big( {{{\hat u}_i},{{\hat d}_i}} \Big)$, $\hat L_i^T = \Big( {{{\hat \nu}_i},{{\hat e}_i}} \Big)$ are $SU(2)$ doublet superfields. $\hat d_j^c$, $\hat u_j^c$ and $\hat e_j^c$ represent the singlet down-type quark, up-type quark and lepton superfields, respectively. Additionally, $Y$, $\lambda$ and $\kappa$ are dimensionless matrices, a vector and a totally symmetric tensor. $a, b=1,2$ are SU(2) indices and $i,j,k=1,\;2,\;3$ are generation indices. In the Eq.~(\ref{super-w}), the first three terms are the same as those of the MSSM. Once the electroweak symmetry is broken (EWSB), the next two terms can generate the effective bilinear terms $\epsilon _{ab} \varepsilon_i \hat H_u^b\hat L_i^a$ and $\epsilon _{ab} \mu \hat H_d^a\hat H_u^b$, with $\varepsilon_i= Y_{\nu _{ij}} \left\langle {\tilde \nu _j^c} \right\rangle$ and $\mu  = {\lambda _i}\left\langle {\tilde \nu _i^c} \right\rangle$. The last two terms explicitly violate lepton number and R-parity, and the last term can generate the effective Majorana masses for neutrinos at the electroweak scale. In this paper, the summation convention is implied on repeated indices.

In the $\mu\nu$SSM, the general soft SUSY-breaking terms are given as
\begin{eqnarray}
&&- \mathcal{L}_{soft}\:=\:m_{{{\tilde Q}_{ij}}}^{\rm{2}}\tilde Q{_i^{a\ast}}\tilde Q_j^a
+ m_{\tilde u_{ij}^c}^{\rm{2}}\tilde u{_i^{c\ast}}\tilde u_j^c + m_{\tilde d_{ij}^c}^2\tilde d{_i^{c\ast}}\tilde d_j^c
+ m_{{{\tilde L}_{ij}}}^2\tilde L_i^{a\ast}\tilde L_j^a  \nonumber\\
&&\hspace{1.8cm} + \: m_{\tilde e_{ij}^c}^2\tilde e{_i^{c\ast}}\tilde e_j^c + m_{{H_d}}^{\rm{2}} H_d^{a\ast} H_d^a
+ m_{{H_u}}^2H{_u^{a\ast}}H_u^a + m_{\tilde \nu_{ij}^c}^2\tilde \nu{_i^{c\ast}}\tilde \nu_j^c \nonumber\\
&&\hspace{1.8cm}  + \: \epsilon_{ab}{\left[{{({A_u}{Y_u})}_{ij}}H_u^b\tilde Q_i^a\tilde u_j^c
+ {{({A_d}{Y_d})}_{ij}}H_d^a\tilde Q_i^b\tilde d_j^c + {{({A_e}{Y_e})}_{ij}}H_d^a\tilde L_i^b\tilde e_j^c + {\rm{H.c.}} \right]} \nonumber\\
&&\hspace{1.8cm}  + \left[ {\epsilon _{ab}}{{({A_\nu}{Y_\nu})}_{ij}}H_u^b\tilde L_i^a\tilde \nu_j^c
- {\epsilon _{ab}}{{({A_\lambda }\lambda )}_i}\tilde \nu_i^c H_d^a H_u^b
+ \frac{1}{3}{{({A_\kappa }\kappa )}_{ijk}}\tilde \nu_i^c\tilde \nu_j^c\tilde \nu_k^c + {\rm{H.c.}} \right] \nonumber\\
&&\hspace{1.8cm}  - \: \frac{1}{2}\left({M_3}{{\tilde \lambda }_3}{{\tilde \lambda }_3}
+ {M_2}{{\tilde \lambda }_2}{{\tilde \lambda }_2} + {M_1}{{\tilde \lambda }_1}{{\tilde \lambda }_1} + {\rm{H.c.}} \right).
\end{eqnarray}
Here, the front two lines contain mass-squared terms of squarks, sleptons and Higgses. The next two lines include the trilinear scalar couplings. In the last line, $M_3$, $M_2$ and $M_1$ denote Majorana masses corresponding to gauginos $\hat{\lambda}_3$, $\hat{\lambda}_2$ and $\hat{\lambda}_1$, respectively. In addition to the terms from $\mathcal{L}_{soft}$, the tree-level scalar potential receives the usual D and F term contributions~\cite{ref3}.

Once the electroweak symmetry is spontaneously broken, the neutral scalars develop in general the VEVs:
\begin{eqnarray}
\langle H_d^0 \rangle = \upsilon_d , \qquad \langle H_u^0 \rangle = \upsilon_u , \qquad
\langle \tilde \nu_i \rangle = \upsilon_{\nu_i} , \qquad \langle \tilde \nu_i^c \rangle = \upsilon_{\nu_i^c} .
\end{eqnarray}
One can define the neutral scalars as
\begin{eqnarray}
&&H_d^0=\frac{h_d + i P_d}{\sqrt{2}} + \upsilon_d, \qquad\; \tilde \nu_i = \frac{(\tilde \nu_i)^\Re + i (\tilde \nu_i)^\Im}{\sqrt{2}} + \upsilon_{\nu_i},  \nonumber\\
&&H_u^0=\frac{h_u + i P_u}{\sqrt{2}} + \upsilon_u, \qquad \tilde \nu_i^c = \frac{(\tilde \nu_i^c)^\Re + i (\tilde \nu_i^c)^\Im}{\sqrt{2}} + \upsilon_{\nu_i^c},
\end{eqnarray}
and
\begin{eqnarray}
\tan\beta={\upsilon_u\over\sqrt{\upsilon_d^2+\upsilon_{\nu_i}\upsilon_{\nu_i}}}.
\end{eqnarray}

The $8\times8$  charged scalar mass matrix $M_{S^{\pm}}^2$ contains the massless unphysical Goldstone bosons $G^{\pm}$, which can be written as~\cite{ref-zhang,ref15,ref-zhang-LFV,ref-zhang1}
\begin{eqnarray}
G^{\pm} = {1 \over \sqrt{\upsilon_d^2+\upsilon_u^2+\upsilon_{\nu_i} \upsilon_{\nu_i}}} \Big(\upsilon_d H_d^{\pm} - \upsilon_u {H_u^{\pm}}-\upsilon_{\nu_i}\tilde e_{L_i}^{\pm}\Big).
\end{eqnarray}
In the unitary gauge, the Goldstone bosons $G^{\pm}$ are eaten by $W$-boson, and disappear from the Lagrangian.
Then the mass squared of $W$-boson is
\begin{eqnarray}
m_W^2={e^2\over2s_{_W}^2}\Big(\upsilon_u^2+\upsilon_d^2+\upsilon_{\nu_i} \upsilon_{\nu_i}\Big),
\end{eqnarray}
where $e$ is the electromagnetic coupling constant and $s_{_W}=\sin\theta_{_W}$ with $\theta_{_W}$ is the Weinberg angle.

\section{Rare decay $\bar{B}\rightarrow X_s\gamma$\label{sec3}}

The effective Hamilton for rare decay $\bar{B}\rightarrow X_s\gamma$ at scales $\mu_b={\cal O}(m_b)$ is written as \cite{ref9,ref10,ref11,ref12,ref13,ref14}
\begin{eqnarray}
&&H_{eff}=-\frac{4G_F}{\sqrt{2}}V_{ts}^\ast V_{tb}\sum_i C_i(\mu)\mathcal{O}_i,
\end{eqnarray}
with $G_F$ denoting the Fermi constant and $V_{ij}$ denoting the quark mixing matrix elements. The Wilson coefficients $C_i(\mu)$ play the role of coupling constants at the effective operators $\mathcal{O}_i$. The definitions of those dimension six effective operators are
\begin{eqnarray}
&&{O_1} = {{\bar s}_i}{\gamma _\mu }P_L{c_j}{{\bar c}_j}{\gamma _\mu }P_L{b_i},\nonumber\\
&&{O_2} = {{\bar s}_i}{\gamma _\mu }P_L{c_i}{{\bar c}_j}{\gamma _\mu }P_L{b_j},\nonumber\\
&&{O_3} = {{\bar s}_i}{\gamma _\mu }P_L{b_i}\sum\nolimits_q {{{\bar q}_j}{\gamma _\mu }P_L{q_j}}, \nonumber\\
&&{O_4} = {{\bar s}_i}{\gamma _\mu }P_L{b_j}\sum\nolimits_q {{{\bar q}_j}{\gamma _\mu }P_L{q_i}},\nonumber \\
&&{O_5} = {{\bar s}_i}{\gamma _\mu }P_L{b_i}\sum\nolimits_q {{{\bar q}_j}{\gamma _\mu }P_R{q_j}}, \nonumber\\
&&{O_6} = {{\bar s}_i}{\gamma _\mu }P_L{b_j}\sum\nolimits_q {{{\bar q}_j}{\gamma _\mu }P_R{q_i}},
\end{eqnarray}
where $P_{L,R}=(1 \mp {\gamma _5})/2$, $O_{1,2}$ are the current-current operators and  $O_{3,...,6}$ are the QCD penguin operators. In addition,
$O_{7,\:8}$ and $\widetilde{O}_{7,\:8}$ are the magnetic and chromomagnetic dipole moment operators, which are defined through
\begin{eqnarray}
&&O_7=\frac{e}{16\pi^2}\bar{s}F\cdot\sigma m_b P_R b,\nonumber\\
&&\widetilde{O}_7=\frac{e}{16\pi^2}\bar{s}F\cdot\sigma m_b P_L b,\nonumber\\
&&{O}_8=\frac{g_s}{16\pi^2}\bar{s}G\cdot\sigma m_b P_R b,\nonumber\\
&&\widetilde{O}_8=\frac{g_s}{16\pi^2}\bar{s}G\cdot\sigma m_b P_L b,
\end{eqnarray}
where $F_{\mu\nu}$ and $G_{\mu\nu}=G_{\mu\nu}^aT^a$ are the electromagnetic and strong field strength tensors, $T^a (a = 1,\ldots,8)$ are $\rm{SU}(3)_c$ generators, and $g_s$ represents the strong coupling respectively.

\begin{figure}
\setlength{\unitlength}{1mm}
\centering
\begin{minipage}[c]{0.9\textwidth}
\includegraphics[width=5.1in]{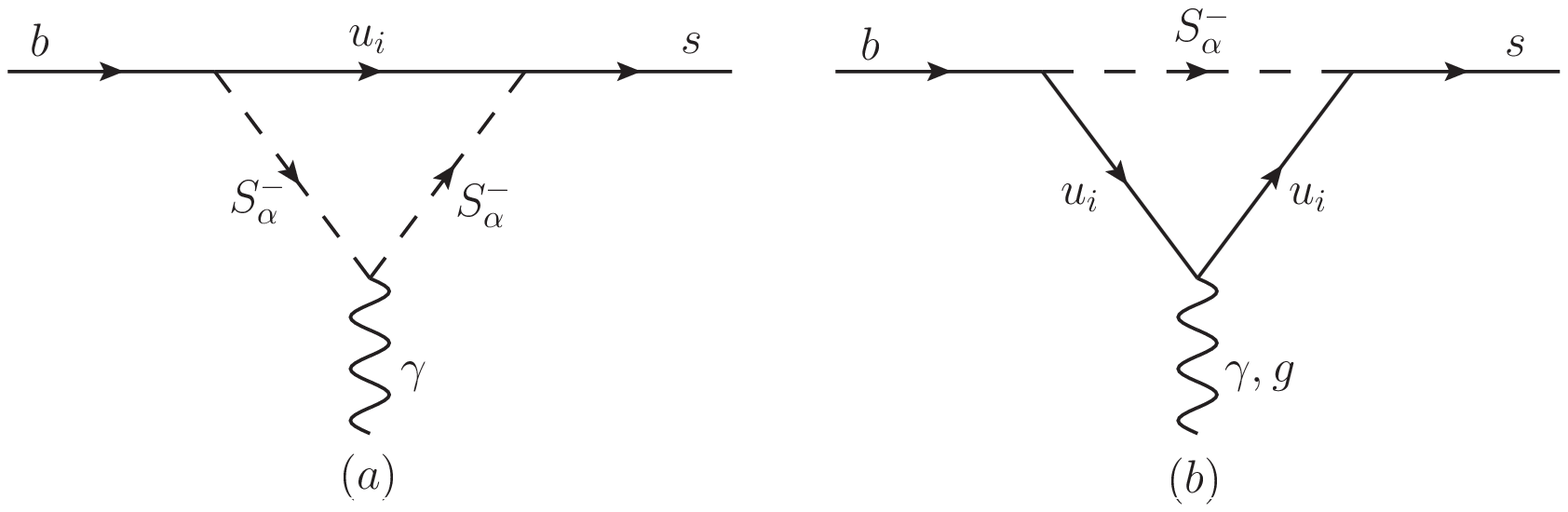}
\end{minipage}
\begin{minipage}[c]{0.9\textwidth}
\includegraphics[width=5.1in]{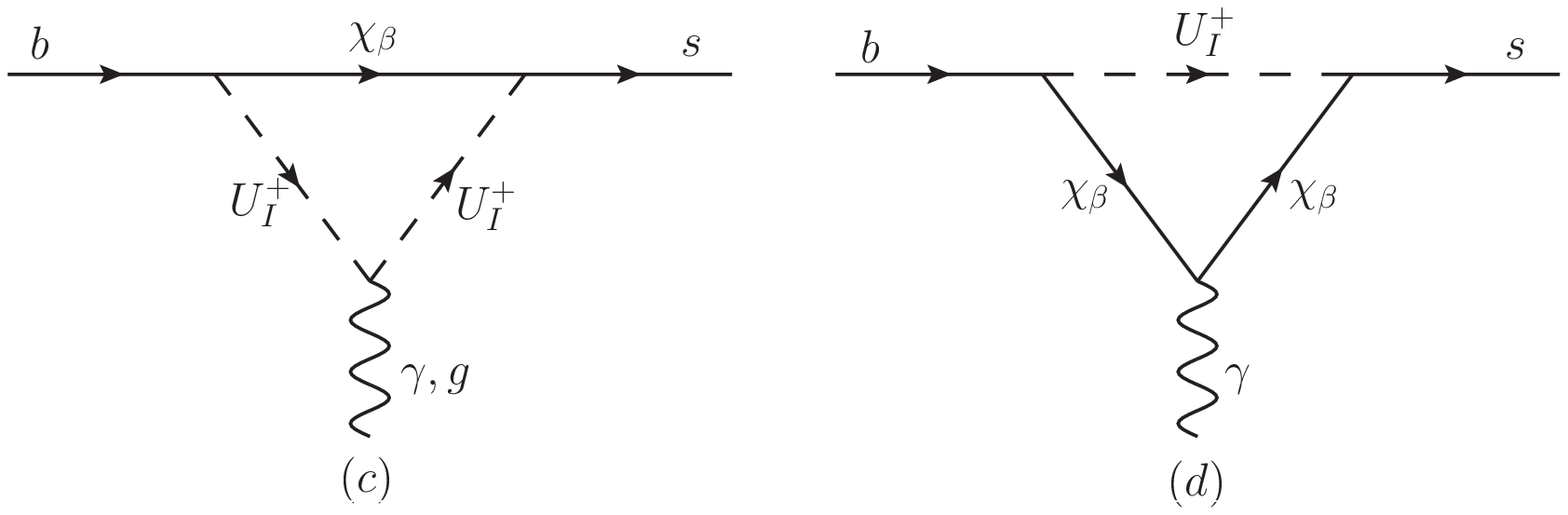}
\end{minipage}
\caption[]{The Feynman diagrams contributing to $\bar{B}\rightarrow X_s\gamma$ from exotic fields in the $\mu \nu$SSM, compared with the SM.}
\label{fig1}
\end{figure}

Compared with the SM, the Feynman diagrams contributing to the process $\bar{B}\rightarrow X_s\gamma$  from exotic fields in the $\mu \nu$SSM are drawn in Fig.~\ref{fig1}, where $S_{\alpha}^-$ ($\alpha=2,\ldots,8$) denote charged scalars, $U_{I}^+$ ($I=1,\ldots,6$) denote up-type squarks, $u_i$ ($i=1,2,3$) denote three generation of up-type quarks and $\chi_{\beta}$ ($\beta=1,\ldots,5$) denote charged fermions.

We could write the Wilson coefficients of the process $b\rightarrow s\gamma$ from the Feynman diagrams in Fig.~\ref{fig1} at the electroweak scale $\mu_{EW}$ as follow:
\begin{eqnarray}
&&C_{7}^{NP}(\mu_{EW})=C_{7\gamma}^{NP}(\mu_{EW})
+\widetilde{C}_{7\gamma}^{NP}(\mu_{EW}),
\end{eqnarray}
where the new physics contributions read
\begin{eqnarray}
&&\widetilde{C}_{7\gamma}^{NP}(\mu_{EW})=\widetilde{C}_{7\gamma a}^{NP}(\mu_{EW})+\widetilde{C}_{7\gamma b}^{NP}(\mu_{EW}) +\widetilde{C}_{7\gamma c}^{NP}(\mu_{EW})+
\widetilde{C}_{7\gamma d}^{NP}(\mu_{EW}),\\
&&\widetilde{C}_{7\gamma a}^{NP}(\mu_{EW})=\sum\limits_{u_{i},S_{\alpha}^-}\frac{s_{_W}^2}{2e^2{V_{ts}^\ast V_{tb}}} \Big\{\frac{1}{2}C_R^{S_\alpha^- {\bar s }u_{i}}C_R^{S_\alpha^{-} {\bar{b}}u_{i}\ast}
  \Big[-{I_3}(x_{{u_{i}}},{x_{{S _{\alpha}^- }}}) + {I_4}(x_{{u_{i}}},{x_{{S _{\alpha}^- }}}) \Big]\nonumber\\
&&\qquad\qquad\qquad\: +\frac{m_{u_{i}}}{m_b} C_L^{S_\alpha^- {\bar s }u_{i}}C_R^{S_\alpha^{-} {\bar{b}}u_{i}\ast}
 \Big[-{I_1}(x_{{u_{i}}},{x_{{S _{\alpha}^- }}})+{I_3}(x_{{u_{i}}},{x_{{S _{\alpha}^- }}}) \Big]\Big\},\\
&&\widetilde{C}_{7\gamma b}^{NP}(\mu_{EW})=\sum\limits_{u_{i},S_{\alpha}^-}\frac{s_{_W}^2}{3e^2{V_{ts}^\ast V_{tb}}}\Big\{\frac{1}{2}C_R^{S_\alpha^- {\bar s }u_{i}}C_R^{S_\alpha^{-} {\bar{b}}u_{i}\ast}
 \Big[-{I_1}(x_{{u_{i}}},{x_{{S _{\alpha}^- }}})+ 2{I_3}(x_{{u_{i}}},{x_{{S _{\alpha}^- }}}) \nonumber\\
&&\qquad\qquad\qquad\:  - {I_4}(x_{{u_{i}}},{x_{{S _{\alpha}^- }}}) \Big] + \frac{m_{u_{i}}}{m_b}C_L^{S_\alpha^- {\bar s }u_{i}}C_R^{S_\alpha^{-} {\bar{b}}u_{i}\ast}
  \Big[ {I_1}(x_{{u_{i}}},{x_{{S _{\alpha}^- }}}) -{I_2}(x_{{u_{i}}},{x_{{S _{\alpha}^- }}}) \nonumber\\
&&\qquad\qquad\qquad\:  - {I_3}(x_{{u_{i}}},{x_{{S _{\alpha}^- }}}) \Big]\Big\},\\
&&\widetilde{C}_{7\gamma c}^{NP}(\mu_{EW})=\sum\limits_{\chi_{\beta},U_{I}^+}\frac{s_{_W}^2}{3e^2{V_{ts}^\ast V_{tb}}}\Big\{\frac{1}{2}C_R^{U_{I}^+ {\bar s }\chi_{\beta}}C_R^{U_{I}^+ {\bar b }\chi_{\beta}\ast}
 \Big[ {I_3}({x_{{\chi_{\beta}}}},{x_{{U _{I}^+ }}}) - {I_4}({x_{{\chi_{\beta}}}},{x_{{U _{I}^+ }}}) \Big]\nonumber\\
&&\qquad\qquad\qquad\: +\frac{m_{\chi_{\beta}}}{m_b}C_L^{U_{I}^+ {\bar s}\chi_{\beta}}C_R^{U_{I}^+ {\bar b}\chi_\beta\ast}
  \Big[ {I_1}({x_{{\chi_{\beta}}}},{x_{{U _{I}^+ }}})- {I_3}({x_{{\chi_{\beta}}}},{x_{{U _{I}^+ }}}) \Big]\Big\},\\
&&\widetilde{C}_{7\gamma d}^{NP}(\mu_{EW})=\sum\limits_{\chi_{\beta},U_{I}^+}\frac{s_{_W}^2}{2e^2{V_{ts}^\ast V_{tb}}}\Big\{\frac{1}{2}C_R^{U_{I}^+ {\bar s }\chi_{\beta}}C_R^{U_{I}^+ {\bar b }\chi_{\beta}\ast}
  \Big[-{I_1}({x_{{\chi_{\beta}}}},{x_{{U _{I}^+ }}})+ 2{I_3}({x_{{\chi_{\beta}}}},{x_{{U _{I}^+ }}}) \nonumber\\
&&\qquad\qquad\qquad\:  -{I_4}({x_{{\chi_{\beta}}}},{x_{{U _{I}^+ }}}) \Big]  +\frac{m_{\chi_{\beta}}}{m_b}C_L^{U_{I}^+ {\bar s}\chi_{\beta}}C_R^{U_{I}^+ {\bar b}\chi_\beta\ast}
 \Big[  {I_1}({x_{{\chi_{\beta}}}},{x_{{U _{I}^+ }}}) -{I_2}({x_{{\chi_{\beta}}}},{x_{{U _{I}^+ }}})
 \nonumber\\
&&\qquad\qquad\qquad\:  -{I_3}({x_{{\chi_{\beta}}}},{x_{{U _{I}^+ }}}) \Big]\Big\},\\
&&C_{7\gamma}^{NP}(\mu_{EW}) = \left.\widetilde{C}_{7\gamma}^{NP}(\mu_{EW})  \right| {_{L\leftrightarrow R}}.
\end{eqnarray}
Here the concrete expressions for coupling coefficients $C_{L,R}$ and form factors $I_{i} $ ($i=1,\ldots,4$) can be found in Appendixes~\ref{appendix-L}--\ref{appendix-integral}. Additionally, $x= {m^2}/{m_W^2}$, where $m$ is the mass for the corresponding particle and $m_W$ is the mass for the $W$-boson.

The Feynman diagrams of the process $b\rightarrow sg$ from exotic fields in the $\mu\nu$SSM compared with the SM are shown in Fig.~\ref{fig1}(b) and Fig.~\ref{fig1}(c). Similarly, the Wilson coefficients of the process $b\rightarrow sg$ at electroweak scale are
\begin{eqnarray}
&&C_{8}^{NP}(\mu_{EW})=C_{8g}^{NP}(\mu_{EW})+\widetilde{C}_{8g}^{NP}(\mu_{EW}),\\
&&\widetilde{C}_{8g}^{NP}(\mu_{EW})=\Big[\widetilde{C}_{7\gamma b}^{NP}(\mu_{EW})+\widetilde{C}_{7\gamma c}^{NP}(\mu_{EW})\Big]/{Q_u},\\
&&C_{8g}^{NP}(\mu_{EW}) = \left.\widetilde{C}_{8g}^{NP}(\mu_{EW})  \right|{ _{L \leftrightarrow R}},
\end{eqnarray}
where $Q_u=2/3$.

In the $\mu\nu{\rm SSM}$, the expression for the branching ratio of $\bar{B}\rightarrow X_s\gamma$ is given as follow
\begin{eqnarray}
&&{\rm{Br}}(\bar{B}\rightarrow X_s\gamma)=R\Big(|C_{7\gamma}(\mu_b)|^2
+N(E_\gamma)\Big)\;,
\end{eqnarray}
where the overall factor $R=2.47\times10^{-3}$, and the nonperturbative contribution
$N(E_\gamma)=(3.6\pm0.6)\times10^{-3}$ \cite{ref13}. $C_{7\gamma}(\mu_b)$ is defined by
\begin{eqnarray}
&&C_{7\gamma}(\mu_b)=C_{7\gamma}^{SM}(\mu_b)+C_{7}^{NP}(\mu_b).
\end{eqnarray}
where we choose the hadron scale $\mu_b=2.5$ GeV and use the SM contribution at NNLO level $C_{7\gamma}^{SM}(\mu_b) = -0.3523$~\cite{ref13,ref14,ref16,ref17-Czakon}. The Wilson coefficients for new physics at the bottom quark scale can be written as \cite{ref19,refGao}
\begin{eqnarray}
&&C_{7}^{NP}(\mu_b)\approx0.5696
C_{7}^{NP}(\mu_{EW})+0.1107 C_{8}^{NP}(\mu_{EW}).
\end{eqnarray}

\section{Numerical analysis\label{sec4}}

There are many free parameters in the SUSY extensions of the SM.  In order to obtain a more transparent numerical results, we adopt the minimal flavor violating (MFV) assumption for some parameters in the $\mu\nu{\rm SSM}$, which assumes
\begin{eqnarray}
&&\lambda _i = \lambda , \qquad {\kappa _{ijk}} = \kappa {\delta _{ij}}{\delta _{jk}}, \quad  {({A_\kappa }\kappa )_{ijk}} ={A_\kappa }\kappa {\delta _{ij}}{\delta _{jk}}, \nonumber\\
&&{({A_\lambda }\lambda )}_i= {A_\lambda }\lambda,\quad {Y_{{\nu _{ij}}}} = {Y_{{\nu _i}}}{\delta _{ij}},\qquad {Y_{{e_{ij}}}} = {Y_{{e_i}}}{\delta _{ij}},
\nonumber\\
&&\upsilon_{\nu_i^c}=\upsilon_{\nu^c},\quad(A_\nu Y_\nu)_{ij}={a_{{\nu_i}}}{\delta _{ij}},\quad{({A_e}{Y_e})_{ij}} = {A_e}{Y_{{e_i}}}{\delta _{ij}},\nonumber\\
&&m_{{{\tilde L}_{ij}}}^2 = m_{\tilde L}^2{\delta _{ij}},\quad
m_{\tilde \nu_{ij}^c}^2 = m_{{{\tilde \nu_i}^c}}^2{\delta _{ij}},\quad
m_{\tilde e_{ij}^c}^2 = m_{{{\tilde e}^c}}^2{\delta _{ij}},\nonumber\\
&&m_{\tilde Q_{ij}}^2 = m_{{{\tilde Q}_i}}^2{\delta _{ij}}, \quad
m_{\tilde u_{ij}^c}^2 = m_{{{{\tilde u}_i}^c}}^2{\delta _{ij}}, \quad
m_{\tilde d_{ij}^c}^2 = m_{{{{\tilde d}_i}^c}}^2{\delta _{ij}},
\label{assumption}
\end{eqnarray}
and one can assume
\begin{eqnarray}
&&(A_u Y_u)_{ij}={A_{u_i}}{Y_{{u_{ij}}}},\quad {Y_{{u _{ij}}}} = {Y_{{u _i}}}{V_{L_{ij}}^u},\nonumber\\
&&(A_d Y_d)_{ij}={A_{d_i}}{Y_{{d_{ij}}}},\quad {Y_{{d_{ij}}}} = {Y_{{d_i}}}{V_{L_{ij}}^d},
\end{eqnarray}
where $V=V_L^u V_L^{d\dag}$ denotes the CKM matrix~\cite{ref23}.
Restrained by the quark and lepton masses, we could have
\begin{eqnarray}
{Y_{{u_i}}} \simeq \frac{{{m_{{u_i}}}}}{{{\upsilon_u}}},\qquad {Y_{{d_i}}} \simeq \frac{{{m_{{d_i}}}}}{{{\upsilon_d}}},\qquad {Y_{{e_i}}} = \frac{{{m_{{l_i}}}}}{{{\upsilon_d}}},
\end{eqnarray}
where $m_{u_i}$, $m_{d_i}$ and $m_{l_i}$ are the up-quark, down-quark and charged lepton masses, respectively, and we choose the values from Ref.~\cite{ref23}.

At the EW scale, the soft masses $m_{\tilde H_d}^2$, $m_{\tilde H_u}^2$ and $m_{\tilde \nu_i^c}^2$ can be derived from the minimization conditions of the tree-level neutral scalar potential, which are given in Refs.~\cite{ref3,ref-zhang}. Ignoring the terms of the second order in $Y_{\nu}$ and assuming $(\upsilon_{\nu_i}^2+\upsilon_d^2-\upsilon_u^2)\approx (\upsilon_d^2-\upsilon_u^2)$, one can solve the minimization conditions of the tree-level neutral scalar potential with respect to $\upsilon_{\nu_i}\:(i=1,2,3)$ as~\cite{ref26}
\begin{eqnarray}
\upsilon_{\nu_i}=\frac{\lambda \upsilon_d (\upsilon_u^2+\upsilon_{\nu^c}^2) - \kappa \upsilon_u \upsilon_{\nu^c}^2}{m_{\tilde L}^2 +{G^2\over 4} (\upsilon_d^2-\upsilon_u^2)} Y_{\nu_i} -\frac{\upsilon_u \upsilon_{\nu^c}}{m_{\tilde L}^2 +{G^2\over 4} (\upsilon_d^2-\upsilon_u^2)}a_{\nu_i},
\label{eq-min}
\end{eqnarray}
where $G^2=g_1^2+g_2^2$ and $g_1 c_{_W} =g_2 s_{_W}=e$.

In the $\mu\nu{\rm SSM}$, the sneutrino sector may appear the tachyons. The masses squared of the tachyons are negative. So, we need analyse the masses of the sneutrinos. The masses of left-handed sneutrinos are basically determined by $m_{\tilde L}$, and the three right-handed sneutrinos are essentially degenerated. The CP-even and CP-odd right-handed sneutrino masses squared can be approximately written as~\cite{ref-zhang1}
\begin{eqnarray}
&&m_{S_{5+i}}^2\approx (A_\kappa+4\kappa\upsilon_{\nu^c})\kappa\upsilon_{\nu^c} +A_\lambda \lambda \upsilon_d \upsilon_u/\upsilon_{\nu^c}-2\lambda^2(\upsilon_d^2+\upsilon_u^2),\\
&&m_{P_{5+i}}^2\approx -3A_\kappa \kappa\upsilon_{\nu^c} +(A_\lambda/\upsilon_{\nu^c}+4\kappa)\lambda \upsilon_d \upsilon_u-2\lambda^2(\upsilon_d^2+\upsilon_u^2).
\end{eqnarray}
Here, the main contribution for the mass squared is the first term as $\kappa$ is large, in the limit of $\upsilon_{\nu^c} \gg \upsilon_{u,d}$. Therefore, we could use the approximate relation
\begin{eqnarray}
-4\kappa\upsilon_{\nu^c}\lesssim A_\kappa \lesssim 0,
\label{tachyon}
\end{eqnarray}
to avoid the tachyons.

Before calculation, the constraints on the parameters of the $\mu\nu{\rm SSM}$ from neutrino experiments should be considered at first. Three flavor neutrinos $\nu_{e,\mu,\tau}$ could mix into three massive neutrinos $\nu_{1,2,3}$ during their flight, and the mixings are described by the Pontecorvo-Maki-Nakagawa-Sakata unitary matrix $U_{_{PMNS}}$ \cite{ref20,ref21}. The experimental observations of the parameters in $U_{_{PMNS}}$ for the normal mass hierarchy show that \cite{ref22}
\begin{eqnarray}
&&\sin^2\theta_{12} =0.302_{-0.012}^{+0.013},\qquad  \Delta m_{21}^2 =7.50_{-0.19}^{+0.18}\times 10^{-5} {\rm eV}^2,  \nonumber\\
&&\sin^2\theta_{23}=0.413_{-0.025}^{+0.037},\qquad  \Delta m_{31}^2 =2.473_{-0.067}^{+0.070}\times 10^{-3} {\rm eV}^2,  \nonumber\\
&&\sin^2 \theta_{13} =0.0227_{-0.0024}^{+0.0023}.
\label{neutrino-oscillation}
\end{eqnarray}

In the $\mu\nu{\rm SSM}$, the three neutrino masses are obtained through a TeV scale seesaw mechanism \cite{ref2,ref26,ref27,ref28,ref29,refneu,refneu1}. Assumed that the charged lepton mass matrix in the flavor basis is in the diagonal form, we parameterize the unitary matrix which diagonalizes the effective light neutrino mass matrix $m_{eff}$ (see Ref.~\cite{ref-zhang}) as \cite{ref30,ref31}
\begin{eqnarray}
{U_\nu} = &&\left( {\begin{array}{*{20}{c}}
   {{c_{12}}{c_{13}}} & {{s_{12}}{c_{13}}} & {{s_{13}}{e^{ - i\delta }}}  \\
   { - {s_{12}}{c_{23}} - {c_{12}}{s_{23}}{s_{13}}{e^{i\delta }}} & {{c_{12}}{c_{23}} - {s_{12}}
   {s_{23}}{s_{13}}{e^{i\delta }}} & {{s_{23}}{c_{13}}}  \\
   {{s_{12}}{s_{23}} - {c_{12}}{c_{23}}{s_{13}}{e^{i\delta }}} & { - {c_{12}}{s_{23}} - {s_{12}}
   {c_{23}}{s_{13}}{e^{i\delta }}} & {{c_{23}}{c_{13}}}  \\
\end{array}} \right)  \nonumber\\
&&\times \: diag(1,{e^{i\frac{{{\alpha _{21}}}}{2}}},{e^{i\frac{{{\alpha _{31}}}}{2}}})\:,
\label{PMNS-matrix}
\end{eqnarray}
where ${c_{_{ij}}} = \cos {\theta _{ij}}$, ${s_{_{ij}}} = \sin {\theta _{ij}}$. In our calculation, the values of $\theta_{ij}$ are obtained from the experimental data in Eq.~(\ref{neutrino-oscillation}), and all CP violating phases $\delta$, $\alpha _{21}$ and $\alpha _{31}$ are set to zero. $U_\nu$ diagonalizes $m_{eff}$ in the following way:
\begin{eqnarray}
U_\nu ^T m_{eff}^T{m_{eff}}{U_\nu} = diag({m_{\nu _1}^2},{m_{\nu _2}^2},{m_{\nu _3}^2}).
\label{eff}
\end{eqnarray}
For the neutrino mass spectrum, we assume it to be normal hierarchical, i.e., ${m_{\nu_1}}{\rm{ < }}{m_{\nu_2}}{\rm{ < }}{m_{\nu_3}}$, and we choose the neutrino mass $m_{\nu_1}=10^{-2}\:{\rm{eV}}$ as input in our numerical analysis, considered that the tiny neutrino masses basically don't affect ${\rm{Br}}(\bar{B}\rightarrow X_s\gamma)$ in the following and limited on neutrino masses from neutrinoless double-$\beta$ decay~\cite{neu-m-limit} and cosmology~\cite{neu-m-limit1}. The other two neutrino masses $m_{\nu_{2,3}}$ can be obtained through the experimental data on the differences of neutrino mass squared in Eq.~(\ref{neutrino-oscillation}).  Then, we can numerically derive $Y_{\nu_i} \sim \mathcal{O}(10^{-7})$ and $a_{\nu_i} \sim \mathcal{O}(-10^{-4}{\rm{GeV}})$ from Eq.~(\ref{eff}). Accordingly, $\upsilon_{\nu_i} \sim \mathcal{O}(10^{-4}{\rm{GeV}})$ through Eq.~(\ref{eq-min}). Due to $\upsilon_{\nu_i}\ll\upsilon_{u,d}$, we can have
\begin{eqnarray}
\tan\beta\simeq \frac{\upsilon_u}{\upsilon_d}.
\end{eqnarray}

Recently, a neutral Higgs with mass around $125\;{\rm GeV}$ reported by ATLAS~\cite{ATLAS} and CMS~\cite{CMS} also contributes a strict constraint on relevant parameter space of the model. The global fit to the ATLAS and CMS Higgs data gives~\cite{mh-AC}:
\begin{eqnarray}
m_{{h}}=125.7\pm0.4\;{\rm GeV}.
\label{M-h}
\end{eqnarray}
Due to the introduction of some new couplings in the superpotential, the SM-like Higgs mass in the $\mu\nu$SSM gets additional contribution at tree-level~\cite{ref3}. For moderate $\tan\beta$ and large mass of the pseudoscalar $M_A$, the SM-like Higgs mass in the $\mu\nu{\rm SSM}$ is approximately given by
\begin{eqnarray}
m_h^2 \simeq m_Z^2 \cos^2 2\beta + \frac{6 \lambda^2 s_{_W}^2 c_{_W}^2}{ e^2} m_Z^2 \sin^2 2\beta+\bigtriangleup m_h^2.
\end{eqnarray}
Compared with the MSSM, the $\mu\nu{\rm SSM}$ gets an additional term $\frac{6 \lambda^2 s_{_W}^2 c_{_W}^2}{ e^2} m_Z^2 \sin^2 2\beta$. Therefore, the SM-like Higgs in the $\mu\nu{\rm SSM}$ can easily account for the mass around $125\,{\rm GeV}$, especially for small $\tan\beta$.
Including two-loop leading-log effects, the main radiative corrections can be given by~\cite{ref-mh-rad1,ref-mh-rad2,ref-mh-rad3}
\begin{eqnarray}
\bigtriangleup m_h^2 = \frac{3m_t^4}{4\pi^2\upsilon^2}\Big[(t+\frac{1}{2}{\tilde X_t})+\frac{1}{16\pi^2} (\frac{3m_t^2}{2\upsilon^2}-32\pi \alpha_3) (t^2+{\tilde X_t}t) \Big],
\end{eqnarray}
with
\begin{eqnarray}
t=\log \frac{M_S^2}{m_t^2},\qquad{\tilde X_t}= \frac{2\tilde A_t^2}{M_S^2} \Big(1-\frac{\tilde A_t^2}{12M_S^2} \Big),
\end{eqnarray}
where $\upsilon=174$ GeV, $M_S =\sqrt{m_{{\tilde t}_1}m_{{\tilde t}_2}}$ with $m_{{\tilde t}_{1,2}}$ being the stop masses, $\alpha_3$ is the strong coupling constant, $\tilde A_t = A_t-\mu\cot\beta$ with $A_t$ denoting the trilinear Higgs-stop coupling and $\mu=3\lambda \upsilon_{\nu^c}$ being the Higgsino mass parameter.

Through the analysis of the parameter space in Ref.~\cite{ref3}, we could choose the reasonable values for some parameters as $\kappa=0.4$, $\lambda=0.2$, $\upsilon_{\nu^c}=1\;{\rm TeV}$ and $m_{\tilde L}=m_{{\tilde e}^c}={A_{e}}=1\;{\rm TeV}$ for simplicity in the following numerical calculation. Through Eq.~(\ref{tachyon}), we could choose ${A_{\kappa}}=-300\;{\rm GeV}$ to avoid the tachyons.  For the Majorana masses of the gauginos, we will imply the approximate GUT relation $M_1=\frac{\alpha_1^2}{\alpha_2^2}M_2\approx 0.5 M_2$ and $M_3=\frac{\alpha_3^2}{\alpha_2^2}M_2\approx 2.7 M_2$. The gluino mass, $m_{{\tilde g}}\approx M_3$, is larger than about $1.2$ TeV from the ATLAS and CMS experimental data~\cite{ATLAS-sg1,ATLAS-sg2,CMS-sg1,CMS-sg2}. So, we conservatively choose $M_2=1\;{\rm TeV}$. The first two generations of squarks are strongly constrained by direct searches at the LHC~\cite{ATLAS-sq1,CMS-sq1}. Therefore, we take $m_{{\tilde Q}_{1,2}}=m_{{\tilde u}^c_{1,2}}=m_{{\tilde d}^c_{1,2}}=2\;{\rm TeV}$. The third generation squark masses are not constrained by the LHC as strongly as the first two generations, and affect the SM-like Higgs mass. So, we could adopt $m_{{\tilde Q}_3}=m_{{\tilde u}^c_3}=m_{{\tilde d}^c_3}=1\;{\rm TeV}$. When the masses of squarks are TeV scale, the contributions to ${\rm{Br}}(\bar{B}\rightarrow X_s\gamma)$ of squarks become small, so we could reasonably use the above choice in the following calculation. For simplicity, we also choose $A_{d_{1,2,3}}=A_{u_{1,2}}=1\;{\rm TeV}$. As a key parameter, $A_{u_3}=A_t$ affects the following numerical calculation. In the limit of $\upsilon_{\nu^c}\gg\upsilon_{u,d}$ \cite{ref-limit-MH}, the charged Higgs mass squared $M_{H^\pm}^2$ in the $\mu\nu$SSM can be formulated as
\begin{eqnarray}
M_{H^\pm}^2\simeq M_A^2+(1-\frac{6s_{_W}^2\lambda^2}{e^2})m_W^2,
\end{eqnarray}
with the neutral pseudoscalar mass squared
\begin{eqnarray}
M_A^2\simeq \frac{6\lambda\upsilon_{\nu^c}(A_\lambda+\kappa\upsilon_{\nu^c})}{\sin2\beta}.
\end{eqnarray}
Considered that $M_{H^\pm}$ also is a key parameter which affects the numerical results, we could take $M_{H^\pm}$ as input to constrain the parameter $A_\lambda$.

\begin{figure}
\setlength{\unitlength}{1mm}
\centering
\includegraphics[width=3.2in]{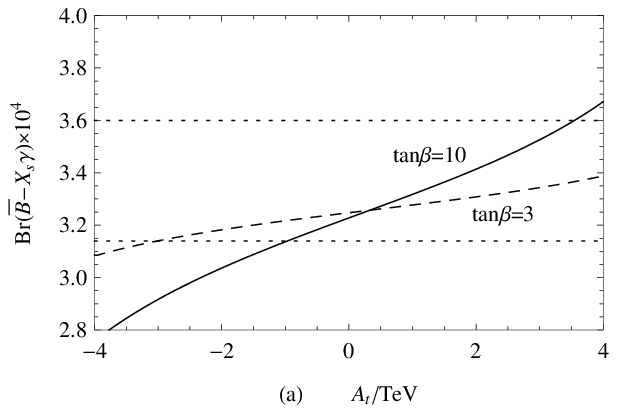}%
\includegraphics[width=3.2in]{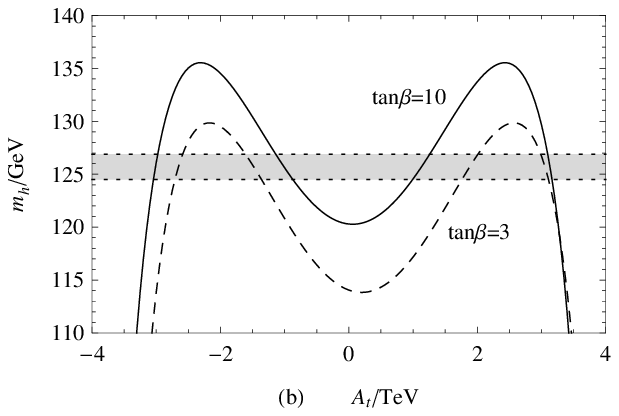}
\caption[]{(a) ${\rm{Br}}(\bar{B}\rightarrow X_s\gamma)$ versus $A_t$ for $\tan \beta=3$ (dashed line) and $\tan \beta=10$ (solid line), when $M_{H^\pm}=1.5$ TeV. The dotted lines represent the experimental $1\sigma$ bounds. (b) The SM-like Higgs mass $m_h$ versus $A_t$ for $\tan \beta=3$ (dashed line) and $\tan \beta=10$ (solid line), where the gray area denotes the experimental $3\sigma$ interval.}
\label{fig-At}
\end{figure}

Similarly to the MSSM and NMSSM~\cite{B-NMSSM}, the new physics contributions to the branching ratio of $\bar{B}\rightarrow X_s\gamma$ in the $\mu\nu{\rm SSM}$ depend essentially on the charged Higgs mass $M_{H^\pm}$, $\tan\beta$ and $A_t$. When $M_{H^\pm}=1.5$ TeV, we plot ${\rm{Br}}(\bar{B}\rightarrow X_s\gamma)$ versus $A_t$ in Fig.~\ref{fig-At}(a), for $\tan \beta=3$ (dashed line) and $\tan \beta=10$ (solid line). The dotted lines represent the experimental $1\sigma$ bounds. The numerical results show that ${\rm{Br}}(\bar{B}\rightarrow X_s\gamma)$ increases with increasing of $A_t$, and the slope of evolution for ${\rm{Br}}(\bar{B}\rightarrow X_s\gamma)$ is big as $\tan \beta$ is large. In Fig.~\ref{fig-At}(a), ${\rm{Br}}(\bar{B}\rightarrow X_s\gamma)$ will be easily below the experimental $1\sigma$ lower bound, when $A_t$ is negative. For positive $A_t$, ${\rm{Br}}(\bar{B}\rightarrow X_s\gamma)$ still can exceed the experimental $1\sigma$ upper bound, as $\tan \beta$ is large enough. So the new physics can give the considerable contributions to ${\rm{Br}}(\bar{B}\rightarrow X_s\gamma)$ for large $\tan\beta$ and $A_t$.

We also need consider the constraint of the SM-like Higgs mass. So in Fig.~\ref{fig-At}(b), we plot the SM-like Higgs mass $m_h$ versus $A_t$ for $\tan \beta=3$ (dashed line) and $\tan \beta=10$ (solid line), where the gray area denotes the experimental $3\sigma$ interval. When $\tan \beta=3$, we require that $A_t$ is about $-2.65$, $-1.5$, $1.9$ or $3.05$ TeV to keep the SM-like Higgs mass around 125 GeV. For $\tan \beta=10$, we need $A_t$ to be about $-3.0$, $-1.0$, $1.1$ or $3.13$ TeV, keeping the SM-like Higgs mass around 125 GeV.

\begin{figure}
\setlength{\unitlength}{1mm}
\centering
\includegraphics[width=3.2in]{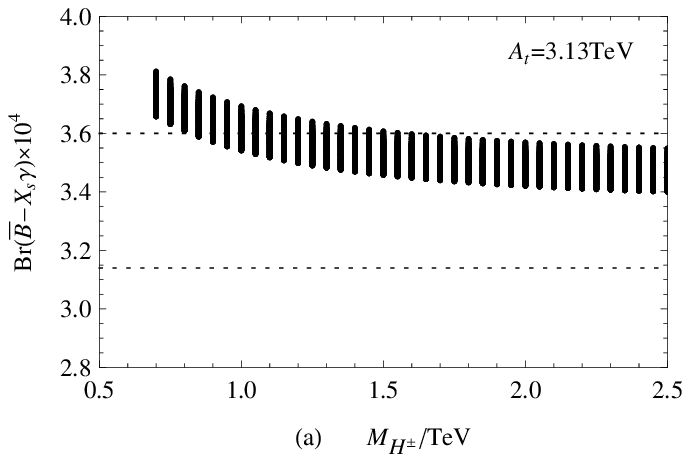}%
\includegraphics[width=3.2in]{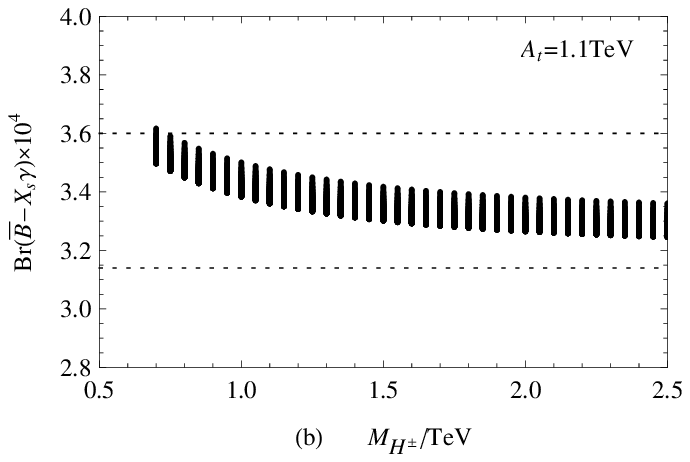}
\includegraphics[width=3.2in]{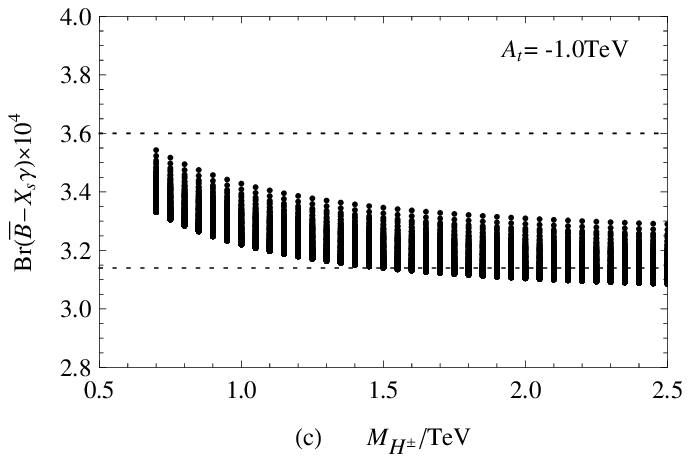}%
\includegraphics[width=3.2in]{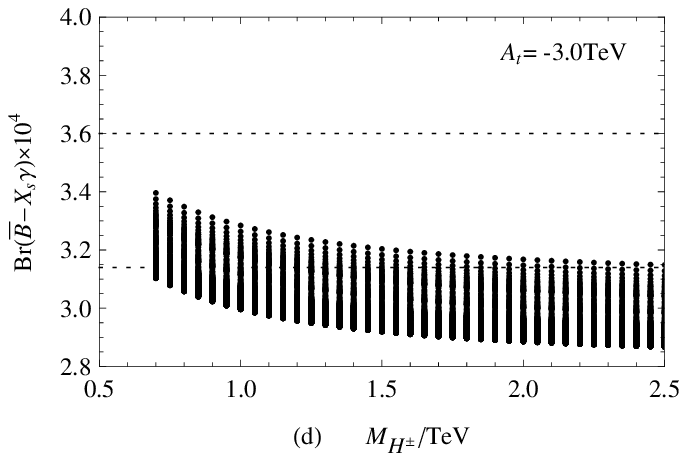}
\caption[]{${\rm{Br}}(\bar{B}\rightarrow X_s\gamma)$ versus $M_{H^\pm}$ for (a) $A_t=3.13$ TeV, (b) $A_t=1.1$ TeV, (c) $A_t=-1.0$ TeV and (d) $A_t=-3.0$ TeV, respectively, when $\tan \beta=10$. Here, we scan over the parameters $\upsilon_{\nu^c}$ and $M_2$ between 0.5 TeV and 1.5 TeV, which step is 0.05 TeV. The horizontal dotted lines represent the experimental $1\sigma$ bounds.}
\label{fig-MHc}
\end{figure}

In large $M_A$ limit, the charged Higgs mass, $M_{H^\pm}\sim M_A \sim M_H$, doesn't affect the SM-like Higgs mass. So, we could choose $A_t=-3.0$, $-1.0$, $1.1$ or $3.13$ TeV, for $\tan \beta=10$, to keep the SM-like Higgs mass around 125 GeV. Then, we draw ${\rm{Br}}(\bar{B}\rightarrow X_s\gamma)$ versus $M_{H^\pm}$ in Fig.~\ref{fig-MHc}, for (a) $A_t=3.13$ TeV, (b) $A_t=1.1$ TeV, (c) $A_t=-1.0$ TeV and (d) $A_t=-3.0$ TeV, respectively, when $\tan \beta=10$. The horizontal dotted lines represent the experimental $1\sigma$ bounds. Here, we scan over the parameters $\upsilon_{\nu^c}$ and $M_2$ between 0.5 TeV and 1.5 TeV, which step is 0.05 TeV. For some $M_{H^\pm}$ and $A_t$, when $\upsilon_{\nu^c}$ and $M_2$ are small, ${\rm{Br}}(\bar{B}\rightarrow X_s\gamma)$ become large. Because the chargino masses are dependent on $\upsilon_{\nu^c}$ and $M_2$, which can give contributions to ${\rm{Br}}(\bar{B}\rightarrow X_s\gamma)$ through chargino-squark loop diagrams in Fig.~\ref{fig1} (c) and (d). Due to constrain the heavy doublet-like Higgs mass $M_H\geq 642$ GeV~\cite{CMS-600,ATLAS-642}, we take the charged Higgs mass $M_{H^\pm}\gtrsim 700$ GeV. The numerical results show that ${\rm{Br}}(\bar{B}\rightarrow X_s\gamma)$ decreases along with increasing of $M_{H^\pm}$, because the contributions from charged Higgs diagrams decay like $1/M_{H^\pm}^4$~\cite{B-NMSSM}. For small $M_{H^\pm}$, the new physics could contribute with large corrections to the branching ratio of $\bar{B}\rightarrow X_s\gamma$. In Fig.~\ref{fig-MHc}, ${\rm{Br}}(\bar{B}\rightarrow X_s\gamma)$ can exceed the experimental $1\sigma$ upper bound for small $M_{H^\pm}$, when $A_t=3.13$ TeV. In addition, ${\rm{Br}}(\bar{B}\rightarrow X_s\gamma)$ can be easily below the experimental $1\sigma$ lower bound for $A_t=-3.0$ TeV, which is excluded by the experimental value at $1\sigma$ level.

\section{Conclusion \label{sec5}}

The flavour changing neutral current process $\bar{B}\rightarrow X_s\gamma$ offers high sensitivity to new physics. In this work, we investigate the branching ratio of the rare decay $\bar{B}\rightarrow X_s\gamma$ in the framework of $\mu\nu$SSM under a minimal flavor violating assumption. Similarly to the MSSM and NMSSM, the new physics contributions to ${\rm{Br}}(\bar{B}\rightarrow X_s\gamma)$ in the $\mu\nu{\rm SSM}$ depend essentially on the charged Higgs mass $M_{H^\pm}$, $\tan\beta$ and $A_t$, because the mixings between charginos and charged leptons in the mass matrix of the $\mu\nu$SSM are suppressed, as well as those between charged Higgses and charged sleptons. Under the constraint of the SM-like Higgs with mass around 125 GeV, the numerical results show that the new physics can fit the experimental data for the rare decay $\bar{B}\rightarrow X_s\gamma$ and further constrain the parameter space. Besides $\bar{B}\rightarrow X_s\gamma$, other $b\rightarrow s$ transitions e.g. $\Delta M_s$,
$S_{J/\psi \phi}$, $B_s\rightarrow \mu^+ \mu^-$ also may give some constraints on relevant parameter space in this model, we will investigate this elsewhere in detail.

\begin{acknowledgments}
\indent\indent
This work has been supported by the National Natural Science Foundation of China (NNSFC)
with Grant No. 11275036, No. 11047002, the open project of State
Key Laboratory of Mathematics-Mechanization with Grant No. Y3KF311CJ1, the Natural
Science Foundation of Hebei province with Grant No. A2013201277, the Natural Science Fund of Hebei University with Grant No. 2011JQ05, No. 2012-242, and the Hundred Excellent Innovation Talents from the Universities and Colleges of Hebei Province with Grant No. BR2-201.
\end{acknowledgments}

\appendix

\section{The interaction Lagrangian\label{appendix-L}}
In the $\mu\nu$SSM, The corresponding interaction Lagrangian of the $\bar{B}\rightarrow X_s\gamma$ process is written as
\begin{eqnarray}
&&\mathcal{L}_{int} = \Big[ S_\alpha^- \bar{d}_i (C_L^{S_\alpha^- \bar{d}_i u_j}{P_L} + C_R^{S_\alpha^- \bar{d}_i u_j}{P_R}) u_j  \nonumber\\
&&\hspace{1.5cm} + \:U_I^+ \bar{d}_i (C_L^{U_I^+ \bar{d}_i{\chi _\alpha}}{P_L} + C_R^{U_I^+ \bar{d}_i{\chi _\alpha}}{P_R}) \chi _\alpha   \Big] + {\rm{H.c.}},
\end{eqnarray}
with $P_{L,R}={(1 \mp {\gamma _5})}/2$,
and the coefficients are
\begin{eqnarray}
&&C_L^{S_\alpha^- \bar{d}_i u_j}=Y_{d_{i}}R_{S^\pm}^{1\alpha} V_{ji}^\ast,\\
&&C_R^{S_\alpha^- \bar{d}_i u_j}=Y_{u_{j}}R_{S^\pm}^{2\alpha} V_{ji}^\ast,\\
&&C_L^{U_I^+ \bar{d}_i{\chi _\alpha}} =  {Y_{d_{i}}} Z_-^{2\alpha}R_u^{jI } V_{ji}^\ast, \\
&&C_R^{U_I^+ \bar{d}_i{\chi _\alpha}} =\Big[- \frac{e}{s_{_W}} Z_+^{1\alpha\ast}R_u^{jI } + {Y_{u_{j}}} Z_+^{2\alpha\ast}R_u^{(3+j)I } \Big] V_{ji}^\ast,
\end{eqnarray}
where $R_{S^\pm}$, $R_{u}$ and $Z_{\mp}$ can be found in Ref.~\cite{ref-zhang}, and $V_{ji}$ denote the quark mixing matrix elements.

\section{Form factors \label{appendix-integral}}
Defining ${x_i} = \frac{{m_i^2}}{{m_W^2}}$, we can have the form factors:
\begin{eqnarray}
&&{I_1}(\textit{x}_1 , x_2 ) =  \frac{{1 + \ln {x_2}}}{{({x_2} - {x_1})}} + \frac{{{x_1}\ln {x_1}}-{{x_2}\ln {x_2}}}{{{{({x_2} - {x_1})}^2}}} ,\\
&&{I_2}(\textit{x}_1 , x_2 ) =  - \frac{{1 + \ln {x_1}}}{{({x_2} - {x_1})}} - \frac{{{x_1}\ln {x_1}}-{{x_2}\ln {x_2}}}{{{{({x_2} - {x_1})}^2}}} ,\\
&&{I_3}(\textit{x}_1 , x_2 ) = \frac{1}{2}\Big[  \frac{{3 + 2\ln {x_2}}}{{({x_2} - {x_1})}} - \frac{{2{x_2} + 4{x_2}\ln {x_2}}}{{{{({x_2} - {x_1})}^2}}} -\frac{{2x_1^2\ln {x_1}}}{{{{({x_2} - {x_1})}^3}}}  +  \frac{{2x_2^2\ln {x_2}}}{{{{({x_2} - {x_1})}^3}}}\Big], \\
&&{I_4}(\textit{x}_1 , x_2 ) = \frac{1}{6} \Big[ \frac{{11 + 6\ln {x_2}}}{{({x_2} - {x_1})}}- \frac{{15{x_2} + 18{x_2}\ln {x_2}}}{{{{({x_2} - {x_1})}^2}}} + \frac{{6x_2^2 + 18x_2^2\ln {x_2}}}{{{{({x_2} - {x_1})}^3}}}  \nonumber\\
&&\hspace{2.4cm} + \: \frac{{6x_1^3\ln {x_1}}-{6x_2^3\ln {x_2}}}{{{{({x_2} - {x_1})}^4}}}  \Big].
\end{eqnarray}


\begin{thebibliography}{99}
\bibitem{refS.C} S. Chen, et al., CLEO Collaboration, Phys. Rev. Lett. 87 (2001) 251807.
\bibitem{refK.A} K. Abe,  et al., BELLE Collaboration, Phys. Lett. B 511  (2001) 151.
\bibitem{refA.L} A. Limosani, et al., BELLE Collaboration, Phys. Rev. Lett. 103  (2009) 241801.
\bibitem{refJ.P} J. P. Lees, et al., BABAR Collaboration, Phys. Rev. Lett. 109 (2012) 191801.
\bibitem{refJ.P1} J. P. Lees, et al., BABAR Collaboration, Phys. Rev. D 86 (2012) 112008.
\bibitem{refJ.PT} J. P. Lees, et al., BABAR Collaboration, Phys. Rev. D 86 (2012) 052012.
\bibitem{refB.A} B. Aubert, et al., BABAR Collaboration, Phys. Rev. D 77  (2008) 051103.
\bibitem{refS.S} S. Stone, PoS ICHEP 2012 (2013) 033.
\bibitem{refmm} M. Misiak, et al., Phys. Rev. Lett. 98 (2007) 022002.
\bibitem{ref18} M. Misiak, M. Steinhauser,  Nucl.~Phys.~B 764 (2007) 62.
\bibitem{refK.C} K. Chetyrkin, M. Misiak, M. Munz, Phys.~Lett.~B~400 (1997) 206.
\bibitem{refC.G} C. Greub, T. Hurth, D. Wyler, Phys.~Rev.~D~54 (1996) 3350.
\bibitem{refK.A1} K. Adel, Y.-P Yao, Phys.~Rev.~D~49 (1994) 4945.
\bibitem{refA.A} A. Ali, C. Greub, Phys.~Lett.~B~361 (1995) 146.
\bibitem{ref2} D.~E.~L\'{o}pez-Fogliani, C.~Mu\~{n}oz,  Phys.~Rev.~Lett. 97 (2006) 041801.
\bibitem{ref3} N.~Escudero, D.~E.~L\'{o}pez-Fogliani, C.~Mu\~{n}oz, R.~Ruiz~de~Austri,  JHEP 0812 (2008) 099.
\bibitem{ref4} J. Fidalgo, D.~E.~L\'{o}pez-Fogliani, C.~Mu\~{n}oz, R.~Ruiz~de~Austri,  JHEP 1110 (2011) 020.
\bibitem{ref5} J.~E.~Kim, H.~P.~Nilles, Phys.~Lett.~B 138 (1984) 150.
\bibitem{ref6} H.~P.~Nilles,  Phys.~Rept. 110  (1984) 1.
\bibitem{ref7} H.~E.~Haber, G.~L.~Kane,  Phys.~Rept. 117  (1985) 75.
\bibitem{ref8} J.~Rosiek, Phys.~Rev.~D 41  (1990) 3464.
\bibitem{ref-zhang} H.-B. Zhang, T.-F. Feng, G.-F. Luo, Z.-F. Ge and S.-M. Zhao, JHEP 1307 (2013) 069 [Erratum ibid. 1310 (2013) 173].
\bibitem{ref15} H.-B.~Zhang, T.-F.~Feng, S.-M.~Zhao, T.-J.~Gao,  Nucl. Phys. B 873 (2013) 300.
\bibitem{ref-zhang-LFV}H.-B.~Zhang, T.-F.~Feng, S.-M.~Zhao, Fei Sun, Int. J. Mod. Phys. A 29 (2014) 1450123.
\bibitem{ref-zhang1}H.-B. Zhang, T.-F. Feng, F. Sun, K.-S. Sun, J.-B. Chen and S.-M. Zhao, Phys. Rev. D 89 (2014) 115007.
\bibitem{ref9} G. Buchalla, A.J. Buras, M.E. Lautenbacher, Rev.~Mod.~Phys.~68 (1996)  1125.
\bibitem{ref10} R. Grigjanis, P.J. O¡¯Donnell, M. Sutherland, H. Navelet, Phys.~Rept.~22 (1993) 93.
\bibitem{ref11} X.-Y Yang, T.-F Feng,  JHEP 1005 (2010) 059.
\bibitem{ref12} L. Lin, T.-F. Feng, F. Sun,  Mod. Phys. Lett. A 24 (2009) 2181.
\bibitem{ref13} A. J. Buras, L. Merlo, E. Stamou,  JHEP 1108 (2011) 124.
\bibitem{ref14} P. Goertz, T. Pfoh, Phys. Rev. D 84 (2011) 095016.
\bibitem{ref16} P. Gambino, M. Misiak,  Nucl.~Phys. B 611 (2001) 338.
\bibitem{ref17-Czakon} M.~Czakon, U.~Haisch, M.~Misiak,  JHEP  0703 (2007)  008.
\bibitem{ref19} A.J. Buras, M. Misiak, M. M\"{u}unz and S. Pokorski, Nucl.~Phys. B 424 (1994) 374.
\bibitem{refGao} T.-J Gao, T.-F Feng, J.-B Chen, Mod. Phys. Lett. A 27 (2012) 1250011.
\bibitem{ref23} J.~Beringer,  et al., Particle Data Group,  Phys.~Rev.~D 86 (2012) 010001.
\bibitem{ref26} P.~Ghosh, S.~Roy, JHEP 0904  (2009) 069.
\bibitem{ref20} B.~Pontecorvo,  Zh. Eksp. Teor. Fiz. 33  (1957) 549.
\bibitem{ref21} Z.~Maki, M.~Nakagawa, S.~Sakata,  Prog. Theor. Phys.  28  (1962) 870.
\bibitem{ref22} M. C. Gonzalez-Garcia, M. Maltoni, J. Salvado, T. Schwetz,  JHEP 1212 (2012) 123.
\bibitem{ref27} P.~Ghosh, P.~Dey, B.~Mukhopadhyaya, S.~Roy,  JHEP 1005  (2010) 087.
\bibitem{ref28} A. Bartl, M. Hirsch, S. Liebler, W. Porodc, A. Vicente, JHEP 0905  (2009) 120.
\bibitem{ref29} J. Fidalgo, D.~E.~L\'{o}pez-Fogliani, C.~Mu\~{n}oz, R.~Ruiz~de~Austri, JHEP 0908 (2009) 105.
\bibitem{refneu} H.-B.~Zhang, T.-F.~Feng, L.-N. Kou, S.-M.~Zhao,  Int. J. Mod. Phys. A 28 (2013) 1350117.
\bibitem{refneu1} H.-B.~Zhang, T.-F.~Feng, Z.-F. Ge, S.-M.~Zhao,  JHEP 1402 (2014) 012.
\bibitem{ref30} S.M.~Bilenky, J.~Hosek, S.T.~Petcov,  Phys.~Lett.~B 94 (1980) 495.
\bibitem{ref31} J.~Schechter, J.W.F.~Valle,  Phys.~Rev.~D 22 (1980) 2227.
\bibitem{neu-m-limit}J. Barea, J. Kotila and F. Iachello, Phys. Rev. Lett. 109 (2012) 042501.
\bibitem{neu-m-limit1}Planck Collaboration, P.A.R. Ade et al, arXiv:1303.5076.
\bibitem{ATLAS} ATLAS Collaboration, Phys. Lett. B 716 (2012) 1.
\bibitem{CMS} CMS Collaboration, Phys. Lett. B 716 (2012) 30.
\bibitem{mh-AC}P.P. Giardino, K. Kannike, I. Masina, M. Raidal and A. Strumia, JHEP 1405 (2014) 046.
\bibitem{ref-mh-rad1}M. Carena, J.R. Espinosa, M. Quir\'{o}s, C.E.M. Wagner, Phys. Lett. B 355 (1995) 209.
\bibitem{ref-mh-rad2}M. Carena, M. Quir\'{o}s, C.E.M. Wagner, Nucl. Phys. B 461 (1996) 407.
\bibitem{ref-mh-rad3}M. Carena, S. Gori, N.R. Shah, C.E.M. Wagner, JHEP 1203 (2012) 014.
\bibitem{ATLAS-sg1}ATLAS Collaboration, Phys. Rev. D 86 (2012) 092002.
\bibitem{ATLAS-sg2}ATLAS Collaboration, JHEP 1310 (2013) 130.
\bibitem{CMS-sg1}CMS Collaboration, JHEP 1301 (2013) 077.
\bibitem{CMS-sg2}CMS Collaboration, JHEP 1307 (2013) 122.
\bibitem{ATLAS-sq1}ATLAS Collaboration, Phys. Rev. D 87 (2013) 012008.
\bibitem{CMS-sq1}CMS Collaboration, JHEP 1210 (2012) 018.
\bibitem{ref-limit-MH}J. Ellis, J.F. Gunion, H.E. Haber, L. Roszkowski, F. Zwirner, Phys. Rev. D 39 (1989) 844.
\bibitem{B-NMSSM}F. Domingo, U. Ellwanger, JHEP 0712 (2007) 090.
\bibitem{CMS-600}CMS Collaboration, Phys. Lett. B 710 (2012) 26.
\bibitem{ATLAS-642}ATLAS Collaboration, ATLAS-CONF-2013-067.




\end{thebibliography}
\end{document}